\documentstyle[12pt,aasms4]{article}
\lefthead{Antiochos, DeVore, \& Klimchuk}
\righthead{Model of Mass Ejections}

\begin{document}

\title{A Model for Solar Coronal Mass Ejections}

\author{S. K. Antiochos}
\affil{E.O. Hulburt Center for Space Research, Naval Research Laboratory,
    Washington, DC, 20375}

\author{C. R. DeVore}
\affil{Center for Computational Physics and Fluid Dynamics, Naval Research Laboratory,
    Washington, DC, 20375}

\and

\author{J. A. Klimchuk}
\affil{E.O. Hulburt Center for Space Research, Naval Research Laboratory,
    Washington, DC, 20375}

\begin{abstract}
We propose a new model for the initiation of a solar coronal mass
ejection (CME).  The model agrees with two properties of CMEs and
eruptive flares that have proved to be very difficult to explain with
previous models.  a) Very low-lying magnetic field lines, down to the
photospheric neutral line, can open toward infinity during an
eruption. b) The eruption is driven solely by magnetic free energy
stored in a closed, sheared arcade; consequently, the magnetic energy
of the closed state is well above that of the post-eruption open
state.  The key new feature of our model is that CMEs occur in
multi-polar topologies, in which reconnection between a sheared arcade
and neighboring flux systems triggers the eruption.  In this
``magnetic breakout'' model, reconnection removes the unsheared field
above the low-lying, sheared core flux near the neutral line,
thereby allowing this core flux to burst open.  We present numerical
simulations which demonstrate that our model can account for the
energy requirements for CMEs. We discuss the implication of the model
for CME/flare prediction.
\end{abstract}


\section{Introduction}

It is now widely recognized that coronal mass ejections (CMEs) are the
most important manifestation of solar activity that drives the space
weather near Earth (\cite{Gosling93,Gosling94}).  CMEs are huge
ejected plasmoids, often with masses $\ga 10^{16}$ g and energies $\ga
10^{32}$ ergs, and frequently subtending more than $60^{\circ}$ in
position angle ({\it e.g.} \cite{Howard85,Hundhausen97}).  Recently,
certain LASCO coronagraph observations from SOHO have been interpreted
as evidence that CMEs can be global disruptions completely circling
the Sun (\cite{Brueckner96,Howard97}). In addition to being the
primary cause of major geomagnetic disturbances, CMEs are also a
fundamental mechanism by which the large-scale corona sheds flux and,
hence, may play a central role in the solar cycle. Therefore, an
understanding of the mechanism for CME initiation has long been a
primary goal of solar physics theory.

There are a number of basic observational constraints that any
theoretical model for CME initiation must satisfy.  Since the plasma
beta in the non-flaring corona is typically small, $\beta \sim
10^{-2}$, gas pressure alone cannot be the driver. Early models for
CMEs proposed that the eruption is driven by explosive flare heating,
but it is now known that many CMEs occur with little detectable
heating, especially those originating from high-latitude quiet
regions. It has also been proposed that CMEs may be due to magnetic
buoyancy effects ({\it e.g.,} \cite{Low94,Wu95,Wolfson97}), but this
would imply CMEs should be associated with large masses of falling
material.  During prominence eruptions, material can sometimes be
observed to fall back onto the chromosphere, but CMEs often occur with
negligible prominence activation and very little evidence for downward
moving plasma. Coronagraph observations usually show all the CME
plasma moving outward, in which case buoyancy is unlikely to be the
driver. These considerations have led most investigators to conclude
that the energy for the eruption must be stored in the magnetic field.

The general topology of a CME appears to be that of an erupting arcade
overlying a photospheric neutral line ({\it e.g.} \cite{Martin97}).  The
magnetic field near CME neutral lines is observed to run almost
parallel to the line, indicating that the field is far from
potential. This magnetic stress is commonly referred to as ``shear'',
and is most likely the source of energy for the eruption.  Note that
the actual type of arcade can range from low-lying kilogauss fields
associated with filament eruptions and flares in active regions to
weak high-latitude fields associated with polar crown filaments.

Since many CMEs are accompanied by the ejection of prominences or
filaments, it appears that the innermost flux of the erupting arcade
opens out to the solar wind. The question of whether all the arcade
flux opens or whether some inner field is left closed is a critical
one for theoretical models
(\cite{Aly91,Sturrock91,Wolfson92,Mikic94}).  All the field that is
observed above the occulting disk of present coronagraphs, $(\ge 1.1
R_{\odot})$, is observed to open, but there may remain a considerable
amount of closed arcade flux below this height. Since low-lying active
region filaments sitting on the chromospheric neutral line are
observed to be ejected, we believe that for some eruptions all the
arcade flux down to the chromospheric neutral line must open.  Others
have argued, however, that the field reconnects back down to a closed
state just as the filament begins to lift, so that the innermost flux
near the chromosphere never actually opens. This issue remains to be
resolved observationally.

As will be discussed in detail below, requiring all the arcade flux to
open imposes an extreme constraint on theories for CMEs. It is much
easier to explain the eruption if a significant fraction of the flux
remains closed.  The key feature of the model presented in this paper
is that it can produce an eruption in which even the innermost arcade
flux at the photospheric neutral line can open to infinity and, yet,
the eruption is purely magnetically driven.

Another observational constraint on theoretical models is that the
stressing of the coronal magnetic field is slow compared to
characteristic time scales in the corona.  Photospheric driving
velocities are typically of order 1 km/s, whereas characteristic
coronal speeds are of order several hundred km/s for sound waves and
several thousand km/s for Alfvenic motions.  Therefore, to a good
approximation, the energy for a CME is pumped into the corona
quasi-statically.

This quasi-static evolution of the corona is controlled by two
processes: the displacement of magnetic footpoints by photospheric and
sub-photospheric flows, and the emergence and submergence of magnetic
flux through the photosphere.  It is not clear which process is
most important for producing the shear, or whether both play
significant roles depending on the type of solar region.  Many CMEs
originate in coronal helmet streamers, which often swell for several
days before the eruption (\cite{Hundhausen97}).  Such swelling often
is attributed to shearing of the magnetic footpoints ({\it e.g.}
\cite{Klimchuk90}), although the emergence of new flux also could
produce this effect. For the model proposed in this paper, it is
irrelevant whether the shear is due to photospheric flows or to flux
emergence. The key requirement is that the stressing is slow. In the
simulations presented below, boundary flows were used to generate the
shear, but we expect the same results if shear emergence is used.  The
major advantage to our boundary conditions is that they are
straightforward to understand and completely physical. As emphasized
by a number of authors ({\it e.g.} \cite{Aly84,Klimchuk89,Wolfson91}),
a low-$\beta$ system like the corona is highly sensitive to boundary
conditions at the base, and the application of inappropriate
conditions can easily lead to erroneous conclusions about the
equilibrium of the coronal field.  By driving the shear with slow
footpoint motions we are assured that any eruption that occurs is
physically valid and not an artifact of peculiar boundary
conditions. Note that the boundary conditions we use are exactly the
ones used by Aly, Sturrock, and others in deriving the energy limit.

The final constraint on CME theories is that the corona has no upper
boundary.  Field lines are allowed to expand to infinity if it is
energetically favorable for them to do so. This seems like a trivially
obvious statement, but it has profound implications for the possible
evolution of the magnetic field.

Taken together, the constraints of an open corona and of quasi-static
stressing make it very difficult to understand how violent eruptions
can occur on the Sun. Intuitively, one would expect the field to
respond to slow photospheric driving by simply expanding outward to
infinity rather than by undergoing a violent eruption.  This intuition
is confirmed by a number of numerical simulations which have shown
that sheared 2.5D force-free bipolar arcades in Cartesian geometry do
not erupt as long as physically appropriate boundary conditions are
assumed (\cite{Klimchuk89,Finn90,Wolfson91}).  There is also no
evidence for eruption in 2.5D bipolar arcades in spherical geometry
(\cite{Roumeliotis94}). The sheared field simply expands outward,
pushing any overlying flux outward as well, and approaches the open
state asymptotically with increasing shear (\cite{Sturrock95}).
Eruptions can be obtained if significant resistivity is present
(\cite{Biskamp89,Steinolfson91,Inhester92,Mikic94}), but contrary to
observations, the field opens only partially and disconnected
plasmoids form in the corona.

The basic problem encountered by all the numerical models to date is
that, irrespective of the magnitude of the shear, the closed
configurations that the models compute contain less magnetic energy
than the fully open state. In fact, Aly and Sturrock have claimed, on
the basis of mathematical arguments, that the maximum energy state of
any force-free field is the fully open field
(\cite{Aly84,Aly91,Sturrock91}).  Although this claim has not been
proved completely rigorously, it is in agreement with every numerical
simulation (including the ones in this paper) and, hence, seems very
likely to be correct.  In order to obtain a CME, however, we need more
energy in the pre-eruption sheared state than in the final open state,
since energy is needed not only to open the field, but also to
accelerate and lift the plasma against gravity. The energy going into
the plasma can be very large, of order $10^{32}$ ergs or more; hence,
the energy in the pre-eruption field must be well above the energy of
the fully open state -- a result which appears to be in direct
conflict with the Aly-Sturrock limit.

\section{The Theoretical Model}

We propose in this paper a CME model that can resolve the apparent
contradiction between the Aly-Sturrock energy limit and CME
observations.  Let us first clarify why Aly-Sturrock, in fact,
requires that previous models be in conflict with observations.  Since
a number of observations indicate that CMEs can occur over extremely
long neutral lines, sometimes apparently circling the whole Sun
(\cite{Brueckner96}), and since most CME and/or flare associated
activity appears to occur along a single neutral line on the
photosphere, models for CMEs and eruptive flares generally assume a
2.5D bipolar magnetic geometry with either translational or azimuthal
symmetry. Hence, the magnetic system consists of a single coronal
arcade ({\it e.g.} \cite{Roumeliotis94,Mikic94}).

The important point is that, for a single arcade system, the
observational requirement that the innermost flux near the neutral
line open up requires that {\it all} the flux in the system open. But
this is forbidden by the Aly-Sturrock limit, at least, for a purely
magnetically-driven eruption, because no sheared field state can have
enough energy to open up all the flux. The only way to obtain an
eruption in these models is by the formation of closed plasmoids in
the corona, so that some of the flux is disconnected from the
photosphere. The Aly-Sturrock arguments do not apply to topologies
with disconnected flux. But in the real 3D corona, reconnection will
not result in the formation of disconnected flux from the
photosphere. In 3D the Aly-Sturrock energy limit will apply and is
likely to suppress any eruption of a single arcade even in the
presence of reconnection, as in the so-called tether-cutting model
(e.g. \cite{Sturrock89,Moore92}). Furthermore, it is very difficult to
see how 3D reconnection in a single arcade can lead to the opening of
the innermost flux near the neutral line.  We conclude that any
single arcade model is doomed to failure.

Let us consider, instead, a multi-flux system such as the field shown
Figure 1a. In this case we have three neutral lines on the photosphere
and four distinct flux systems: a central arcade straddling the
equator (blue field lines), two arcades associated with the neutral
lines at $ \pm 45^{\circ}$ (green field lines), and a polar flux
system overlying the three arcades (red lines). Note that there are
two separatrix surfaces which define the boundaries between the
various flux systems, and a null point on the 
equatorial plane at the intersection of the separatrices.

Suppose that, as a result of large shear concentrated at the
equatorial neutral line, a CME occurs and the innermost flux of the
central arcade opens up. The difference now is that, even if the
system is strictly 2.5D, the innermost flux can open down to the
neutral line at the equator without opening all the flux in the system
and without the formation of disconnected flux.  Clearly, there is no
reason for a shear that occurs only near the equator to affect the
high-latitude (green) arcades; therefore, their flux should remain
closed throughout the evolution.  But this is not significant
energetically, since the green arcades are not expected to play an
important role in constraining the eruption. The key point is that
much of the flux of the central (blue) and overlying (red) systems
also can remain closed during eruption. If reconnection between red
and blue field lines occurs at the null point, this reconnection will
result in the transfer of flux from the blue and red systems to the
side (green) arcades. Such a flux transfer allows unsheared blue and
red flux to get out of the way of the erupting sheared core flux, but
still remain closed.  Note that this process is possible only in a
multi-flux system. In a single arcade system all the unsheared flux is
forced to open because it basically has nowhere else to go.

Flux transfer in a multi-polar system also allows us to understand how
a large energy excess sufficient to power a CME can be built up in a
closed sheared field and, yet, be compatible with the Aly-Sturrock
limit. The key point is that the ``fully'' open state is not unique in
a multi-polar system.  To illustrate, let $E_{max}$ denote the energy
of that state in which all the flux initially in the central and
overlying arcades (blue and red) opens, but the green flux remain
closed.  When the shear is confined only to the arcade at the equator,
$E_{max}$ is the appropriate energy limit rather than the larger
energy corresponding to all the flux, including green, being open.
Note that the state with energy $E_{max}$ can be reached by a purely
ideal evolution in which the sheared core field expands slowly outward
pushing all the overlying flux out to infinity, with no reconnection
at the null. The side flux is distorted by the expanding field, but
does not open. The final state has a current sheet at the equator
reaching down to the photosphere, and a current sheet at the top of
each side arcade, so that each green side arcade becomes a helmet
streamer.

We expect that no sheared closed configuration can have energy above
$E_{max}$, so if it were the only state accessible, there would be no
eruption, just as in the single arcade case. If we allow reconnection
at the null, however, then other open states with lower energy become
accessible.  For example, if the shear is confined to a small
latitudinal band $\pm \Theta$ near the equator and well inside the
separatrix of the blue arcade, then consider the state in which the
flux inside $\pm \Theta$ opens, while as much as possible of the rest
of the flux in the system is closed. (It may not be possible for all
the unsheared blue flux to be closed, because there may not be
sufficient red flux for the blue to reconnect with.) Let this state,
which has the minimal amount of open flux, have energy $E_{min}$.  The
open state $E_{min}$ looks very similar to the state $E_{max}$, except
that the distribution of flux between the open and closed systems is
different.  We can consider $E_{max}$ and $E_{min}$ to be the
maximally and minimally open states respectively. The important point
is that the magnitude of $E_{min}$ can be much less than $E_{max}$,
because the state $E_{min}$ may contain much less open field.

We propose that the energy for CMEs and eruptive flares is due to this
difference between $E_{max}$ and $E_{min}$. If the reconnection at the
null is slow, then the energy in the sheared closed configuration can
rise well above $E_{min}$. It can never go above $E_{max}$, but
depending on the ratio of sheared flux to unsheared flux in the
central arcade, the energy of the sheared closed field could be much
larger than $E_{min}$ so there could be a large amount of energy
available for eruption.

In this model the unsheared blue and red fluxes act to confine the
expansion of the sheared core field. We expect that a CME (or
eruptive flare) occurs when reconnection at the null weakens this
confinement sufficiently, so that the sheared field starts to expand
outward rapidly, driving ever faster reconnection at the null and
``breaking out'' to infinity.  Note that this magnetic ``breakout''
model naturally implies explosive-type behavior.

A crucial issue for our model is the rate of reconnection at the
null. If this rate is fast, faster than the rate at which the system
is being sheared (either by photospheric flows or flux emergence),
then the magnetic energy always stays below $E_{min}$ and no eruption
can occur. We expect, however, that the reconnection will be slow at
first because the shear occurs only far from the separatices.  Under
these circumstances a quasi-static evolution should be possible, so
that only weak currents develop near the null and separatrices. If so,
there should be negligible reconnection, and the energy in the system
may rise well above $E_{min}$. In the next section we present
numerical simulations which confirm this hypothesis.

\section{Numerical Simulations}

\subsection{Force-Free Field Calculations}

In order to test the free energy issue discussed above, we calculated
the energy of a sheared multipolar field using two procedures. First,
we calculated the energy $E_{min}$ by determining the minimum-energy
force-free field for a given distribution of shear at the photosphere.
The use of a force-free calculation needs some clarification.  Since
the field has a null point, the true evolution cannot be force-free
because any finite gas pressure will dominate near the null. Hence,
the force-free simulations do not determine a physical evolution for
the system. We use the force-free code only as a convenient method for
calculating the equilibrium state corresponding to the minimum
magnetic energy for a given shear on the field and in the limit of
vanishing gas pressure.

The important feature of the force-free calculations is that they
include the effects of reconnection.  The numerical algorithm that the
code uses to solve the force-free equations is basically iterative
relaxation (\cite{Yang86}).  Given the positions of all the field line
footpoints on the boundary, and some initial guess at the solution in
the interior, the code iterates the solution monotonically towards the
minumum energy state. However, the code uses a multi-grid procedure,
so that the relaxation is first performed on very coarse grids with
poor resolution and, consequently, high numerical diffusion.  This
means that the iteration process allows for reconnection, in that flux
will transfer rapidly from one flux system to the other as long as
this transfer is consistent with the boundary conditions and results
in a decrease in the magnetic energy. Note that there is no real
diffusion in the system, however, because there is no slippage of the
footpoints at the solar surface. The shear is strictly enforced by the
boundary conditions, but the field is free to reconnect at will across
the null in order to reach its absolute minimum energy state.

The code solves the force free equations,
\begin{equation}
(\nabla \times \vec{B}) \times \vec{B} = 0,
\end{equation}
using an Euler potential representation for the field (\cite{Yang86}).
Azimuthal symmety is assumed, so that
\begin{equation}
\vec{B} = \nabla\alpha(r,\theta) \times \nabla \left(\phi - \gamma(r,\theta)\right).
\end{equation}
The advantage of the Euler representation is that the photospheric shear 
is specified directly by the value of
$\gamma$ at the photosphere, $r = 1$. By fixing $\gamma$ there, we fix the
shear irrespective of how the field changes in the interior.

For the initial, unstressed (potential) field (Figure 1a), we assumed a 
multipolar flux distribution given by,
\begin{equation}
\alpha(r,\theta) = \frac{\sin^2\theta}{r} + 
\frac{(3 + 5\cos 2\theta)\sin^2\theta}{2 r^3}.
\end{equation}
The field consists of a dipolar component which 
dominates at large $r$, and an octopolar component at the surface. In the
middle corona, $r\sim 2$, the field appears quadrupolar with an 
X-type null point on the equatorial plane.
Due to the 2.5D symmetry of our system, the normal flux at the
photosphere is unchanged by the shear and, hence, the value of $\alpha$ at the
lower boundary is constant:
$\alpha(1,\theta) = (5/4) \sin^2 2\theta$. Note that the flux maximum
$\alpha = 1.25$ occurs at the neutral line at $\theta = \pi/4$.

From Eqtns (2) and (3) we find that,
\begin{equation}
B_r(r,\theta) = \frac{1}{r^2 \sin \theta} \frac{\partial \alpha}
                {\partial \theta}.
\end{equation}
so that the normal field at the photosphere is given by $B_r(1,\theta)
= 10 \cos \theta \cos 2\theta$, which implies that there are three
neutral lines located at $\theta = \pi/4, \, \pi/2,$ and $3\pi/4$. The
position of the null point on the equatorial plane can be determined
from the requirement that $B_{\theta}$ vanish there. From (2) and (3)
we find that $B_{\theta}(r,\pi/2) = r^{-3} - 3r^{-5}$, consequently
the null is located at $r = \sqrt{3}$. The value of the flux function
at the null, $\alpha(\sqrt{3},\pi/2) = 2/(3\sqrt{3}) = .385$, from
which we find that the separatrices intersect the northern hemisphere
at $\theta = .2941$ and 1.277. Note that we have chosen the form of
$\alpha$ so that positive flux in each hemisphere exactly balances the
negative flux there, while $\alpha$ vanishes at the poles and
equator. This means that there is exactly enough red flux to reconnect
with all the blue flux, if such reconnection were energetically favored.

The imposed shear is assumed to be
antisymmetric about the equator, and confined to a small 
latitudinal band there; i.e.,
\begin{equation}
\gamma(1,\theta) = \left\{ \begin{array}{ll}
\chi\, C  (\psi^2 - \Theta^2)^2 \sin \psi,  & 
\mbox{for $\psi < \Theta$} \\
0 & \mbox{for $\psi \ge \Theta$}, 
\end{array}
\right.
\end{equation}
where $\psi = (\pi/2 - \theta)$ is the solar latitude,
$\Theta = \pi/15$ is the assumed latitudinal extent of the shear layer, 
and $C = 8.68252 \times 10^3$ is a normalization constant 
chosen so that
$\gamma = \chi$ at the latitude of maximum shear, $\psi = .094$. 
The form of
$\gamma(1,\theta)/ \chi$ is shown in Figure 2. Note that the
boundary of the shear layer occurs at $\theta = 13\pi/30$, where
$\alpha = .207$, whereas the value of $\alpha$ at the
boundary of the blue arcade (the separatrix) is .385. Hence, 
only about half the flux of the central arcade is sheared.

We solved the force-free equations above in the domain $(1 \le r \le
100, \, 0 \le \theta \le \pi/2)$ using a $512 \times 512$ non-uniform
grid.  Due to the symmetry of the system about the equatorial
plane, only one hemisphere needed to be calculated. The grid was selected
to enhance the resolution near the solar surface and near the
equator, in particular, the grid points were spaced uniformly in $x$
and $y$, where $x = \ln r$ and $y = \exp(6\,\theta/\pi)$.  The
boundary conditions at the pole $\theta = 0$ and the equator $\theta =
\pi/2$ were set by symmetry. At the inner boundary $r = 1$, the shear
profile $\gamma(1,\theta)$ was specified using the form above, and the
flux distribution $\alpha(1,\theta)$ was fixed to its initial
value. At the outer boundary $r=100$ both $\alpha(100,\theta)$ and
$\gamma(100,\theta)$ were fixed to their initial value. 
This meant that there was no shear at the outer boundary,
$\gamma(100,\theta) = 0.$, which is a good assumption, because
the sheared field lines were initially very far from the outer
boundary.  However, fixing $\alpha$ at the outer boundary means that
no flux is allowed to exit the system. This is not valid since as
the inner core field expands, it will push on the outer field lines so
that even the flux very far from the solar surface will move outward.
The amount of flux at $r=100$ is very small, however, and the energy
associated with this flux is so small that the outer boundary
conditions have a negligible effect on the structure and energetics of
the system, even for shears much larger than the values we discuss
below.

The results of the force-free field calculation for three shear
values, $\chi = \pi/8$, $\pi/4$, and $\pi/2$, are shown in Figures 1b,
c, and d. The field lines in all these figures were traced beginning
at exactly the same footpoint positions as in Figure 1a, in which
there are 4 footpoints in the red system, 3 green footpoints, 3
light-blue (unsheared), and 3 dark-blue (sheared).  The color of the
field lines in Figures 1b - 1d corresponds to the initial color of the
footpoints, irrespective of whether reconnection has occurred.  Even
for a low shear of $\pi/8$, it is evident that some reconnection has
taken place since two of the red and two of the light-blue field lines
have now joined the green system.  It should be noted that, although
the maximum shear is $\pi/8$, the average over the shear layer is only
about half this value. For $\chi=\pi/2$ almost all the blue flux has
reconnected with the red and has expanded outward far from its initial
position. Therefore, we expect the energy of the $\pi/2$ sheared state
to be close to that of the open state, $E_{min}$. In fact, we have
continued the force-free calculations to shears twice this value and
find almost no increase in the magnetic energy. It can be seen in
Figure 3 that the energy of the force-free field appears to saturate
at a maximum value of approximately 6\% above the initial potential
field energy, i.e. $E_{min} = 1.06 E_{pot}$. Note that the 6\% figure
refers to the total energy of the initial field, including all the
unsheared arcades. If we consider only the energy initially in the
central, blue arcade, the relative energy increase is over an order of
magnitude larger.

An issue that requires clarification is how the force-free
calculations determine a sequence of equilibrium states leading to the
state $E_{min}$.  The code finds the absolute minimum energy state for
a given shear at the base, assuming that reconnection occurs freely at
the null. For small shear only the unsheared light-blue field lines
reconnect, as in Fig. 1b, but as the shear increases the outermost
sheared field line eventually reaches the null. It is important to
emphasize that the sheared field lines (dark-blue) also reconnect
freely.  This is evident from Figs. 1c and 1d. The dark-blue lines
have reconnected and now have one of their footpoints near the
pole. Although these lines have reconnected, they have not lost their
shear. The footpoint displacement $\gamma$ is strictly maintained at
the photosphere by the boundary conditions. This is also physically
valid.  As long as the system has small resistivity, the reconnection
will occur in a region of negligible volume (a current sheet), and the
field reconnects with negligible diffusion.  In this case the
reconnected field lines maintain their footpoint displacement
(\cite{Karpen96}).  Note also that all field lines with a footpoint in
the shear zone, even if the field line has joined the green system,
continue to expand outward toward the open configuration as the shear
increases.  In the limit of infinite shear, the field achieves the
open state $E_{min}$ in which all the field lines in the shear zone
(by now they have all become part of the green systems due to
reconnection) and any left over red field lines open.

\subsection{MHD Calculations}

The results of our force-free calculations determined the magnitude of
$E_{min}$. The key question is whether a sheared coronal field will
evolve to a state with energy significantly above this value, or
whether reconnection at the null keeps the energy below $E_{min}$.  In
order to calculate the physical evolution for the field, we performed
a fully time-dependent simulation using a 2.5D ideal MHD code in
spherical coordinates developed by DeVore (1991). The code uses a
multi-dimensional FCT algorithm for transporting the conserved
quantities, and maintains the divergence free condition on the
magnetic field to machine accuracy.

The code solves the standard ideal MHD equations appropriate for the
solar corona:

\begin{equation}
\frac{\partial \rho}{\partial t} + \nabla \cdot (\rho \vec{v}) = 0
\end{equation}
\begin{equation}
\frac{\partial}{\partial t}(\rho \vec{v}) + \nabla \cdot (\rho
\vec{v} \vec{v}) + \nabla P = \frac{1}{4\pi}(\nabla\times\vec{B})
\times\vec{B} - \rho g_{\odot} R^2_{\odot} \frac{\vec{r}}{r^3} 
\end{equation}
\begin{equation}
\frac{\partial U}{\partial t} + \nabla \cdot (U \vec{v}) +
P \nabla \cdot \vec{v} = 0
\end{equation}
\begin{equation}
\frac{\partial \vec{B}}{\partial t} = \nabla \times 
(\vec{v} \times \vec{B})
\end{equation}
where $\rho$ is mass density, $\vec{v}$ is velocity, 
$P$ is gas pressure, $\vec{B}$ is magnetic induction, $g_{\odot}$
is the solar surface gravity ($2.75\times 10^4$ cm/s$^2$), $R_{\odot}$
is the solar radius ($7.0 \times 10^{10}$cm), and $U$ is the 
internal energy ($U = 3P/2$). 

We use exactly the same initial magnetic field and shear profile as in
the force-free calculations, but now we must also specify the
initial plasma distribution. The plasma was taken to be
in hydrostatic equilibrium, with initial temperature and
density given by
\begin{equation}
T(r) = \frac{2\times 10^6}{r^7} {\rm K}, \quad {\rm and} \quad
n(r) = \frac{2\times 10^8}{r} \rm{cm}^{-3}.
\end{equation}
Since the field drops off rapidly with radius, these forms for $T$ and
$n$ were chosen in order to keep the plasma $\beta \equiv 8\pi P/B^2$
from becoming too large at large distances from the surface.  Although
the values of $\beta$ in our simulation are not as low as those in the
real corona, they are definitely less than unity near the bottom
boundary.  The values for $T$ and $n$ above imply an initial plasma
pressure at the solar surface of $5.5 \times 10^{-2}$ergs/cm$^3$. The
field strength at the surface ranges from a value of $B = B_r = 10$ at
the pole, so that $\beta = .014$ there, to $B = B_{\theta} = -2$ at
the equator where $\beta = .35$. Once we begin shearing, the value of
$\beta$ drops much lower. Of course the system is always high-beta
near the null point, so it deviates substantially from the force-free
case there.

We use the same non-uniform grid as in the force-free calculations,
except that the outer boundary is placed at $r=10$ rather than at
$r=100$. As will be shown below, the field does not expand outward as
much in the MHD case, so there is less influence from the outer
boundary.  Furthermore, we assume open boundary conditions there, so
that flux and mass are free to move past the outer boundary. The use
of a smaller domain allows us to have higher spatial resolution, which
turns out to be the limiting factor in the MHD simulations.  The
boundary conditions at the polar axis and the equatorial plane are
determined, as before, by symmetry considerations. At the inner
boundary, the photosphere, we assume a line-tied impenetrable surface
with an imposed azimuthal velocity given by the shear profile above
and with a sinusoidal temporal profile.

A critical parameter is the magnetic Reynolds number $R_M$. Our
fully-2D FCT transport algorithm uses higher order differencing so
that, as in all finite-difference codes, the effective Reynolds number
is quite high: $R_M >> 10,000$, as long as the magnetic and plasma
structure is well resolved numerically. Of course, these values for
$R_M$ are still small compared to classical solar values. If structure
develops on the scale of the grid spacing, then the effective $R_M$
drops to become of order the number of grid points $\sim 500$. We
conclude, therefore, that our simulation only overestimates
(greatly) the true rate of reconnection that would occur on the
Sun.

We drove the system with a sequence of shear motions applied at the
photospheric boundary. Each shear phase had a sinusoidal time profile
with a period of 25,000 s and a maximum footpoint displacement of
$\pi/8$.  For this period and amplitude the maximum shear velocity at
the photosphere is approximately 10 km/s, whereas the Alfven speed is
approximately 500 km/s in the model corona.  Hence, the evolution was
nearly quasi-static until the very end of the simulation when large
velocities appeared near the null due to reconnection.

The resulting field lines at the end of the first shear phase are
shown in Figure 4a. These field lines are traced from exactly the same
footpoint positions as the lines in Figure 1a.  We note that the field
lines in Fig. 4a have pushed outward as a result of the shear, but
unlike the corresponding force-free case (Fig. 1b) there is no
evidence for reconnection yet in the MHD results. The MHD system does
appear to have achieved a true equilibrium. We let the system relax
for an additional 50,000 s, and the resulting field was virtually
indistinguishable from that in Fig. 4a.  We conclude that there are
at least two magnetostatic equilibria that the system can achieve for
a given shear. One is the force-free state of Fig. 1b, in which the
field outside the sheared region is current-free and the null is a
true X-point with right-angle separatrix lines. This state has the
lowest energy, but it can be reached only with reconnection.  Another
solution is the MHD state of Fig. 4a, in which no reconnection
occurs at the null. In this case the unsheared field cannot be
current-free, so that gas pressure, and perhaps gravity, must play a
role in the force balance. We can infer from Fig. 4a that the
separatrix surfaces are not quite perpendicular to each other, which
indicates there is some weak current near the null. This current,
however, is less than 1\% of the maximum current in the shear region.

Figures 4b, c, and d show the results of the MHD simulation after
three more shear phases. The solution in Fig. 4c corresponds to a
total maximum shear in each hemisphere of $3 \pi/8$. Although no
noticeable reconnection has ocurred, the field at the null becomes
progressively distorted from a right-angle X-type neutral point. The
reason for this is straightforward. To relieve their stress, blue
field lines expand outward towards an open configuration, pushing the
overlying red lines outward as well.  The green field lines, on the
other hand, are unstressed and have no interest in being dragged out
to infinity, so they simply move aside by pulling away from the
neutral point. Consequently, the separatrix lines become more and more
oblique, as blue and red flux push towards each other while green
pulls away.

We can gain some insight into the evolution of the structure near the
null point by considering a simple analytic model.  Curvature effects
can be neglected sufficiently near the null and a locally Cartesian
coordinate system can be used.  Let the origin of this coordinate
system be at the null, the $x$ axis be horizontal (parallel to the
equatorial plane), and the $y$ axis be vertical. If the field near the
null has no shear, then an appropriate form for the potential $\alpha$
(see Equation (2)) there is:
\begin{equation}
\alpha(x,y) = B_0 \left( \frac{y^2}{2l_y} - \frac{x^2}{2l_x}\right),
\end{equation}
where $l_x$ and $l_y$ are two constants that determine the scale of the
gradients in $x$ and $y$ respectively. The separatrix lines are given by
$\alpha = 0$; consequently, if $l_x = l_y$, these lines are perpendicular
to each other, and as we show below, the current vanishes.
From this form for $\alpha$ we derive a magnetic field,
\begin{equation}
\vec{B}=  B_0 \left( \frac{y}{l_y}  \hat x   +  \frac{x}{l_x} \hat y \right), 
\end{equation}
and an electric current density, 
\begin{equation}
\vec{J}=  B_0 \frac{l_y - l_x}{l_xl_y} \hat z.
\end{equation}
Note that
if $l_x=l_y$ then the current vanishes and we recover the right-angle X-type
null point of a potential field. If the scales are not equal, then there is
a finite current in the system which implies a finite 
Lorentz force given by,
\begin{equation}
\vec{J} \times \vec{B} = B_0^2 \frac{l_y-l_x}{l_xl_y}\left(-\frac{x}{l_x}\,\hat x
  + \frac{y}{l_y}\,\hat y \right).
\end{equation}
Assuming that gravity is negligible, then the Lorentz force must be balanced by 
the pressure gradient, which implies that the pressure is given by,
\begin{equation}
P =  P_0 + \frac{B_0^2}{4\pi} \frac{l_y-l_x}{l_xl_y}\left(-\frac{x^2}{2l_x} 
+ \frac{y^2}{2l_y} \right),
\end{equation}
where $P_0$ is the gas pressure at the null.

Initially the field is current-free and $l_x = l_y$, but as the field
is distorted by the outward expanding blue flux, the angle formed by
the separatrix surfaces becomes more and more acute so that $l_x$
becomes smaller than $l_y$. In the limit that $l_x << l_y$, Equations
(12) -(14) above imply that, $\vec{B} \to (B_0 x/l_x)\hat y$, $\vec{J}
\to (B_0/l_x) \hat z$, and $P \to P_0 -(B_0^2/8\pi) (x^2/l_x^2)$.  In
other words, the null deforms to a classical current sheet.  This
trend is clearly evident in Fig. 4.

From our investigations of single dipole fields
(\cite{Roumeliotis94,Sturrock95}) we expect that at large shear, the
red and blue field lines expand outwardly exponentially with
increasing shear and, consequently, the current sheet thins out
exponentially fast. This implies that reconnection eventually must
begin to occur here, and once this reconnection starts it should
accelerate.  Our simulation shows convincing evidence for this type of
evolution.  There is no sign of reconnection at a shear of $3\pi/8$
(Fig. 4c). We let the field relax for an additional 30,000 s after this
shearing phase, and saw no change in the field structure; hence, the
field in Fig. 4c is also a true equilibrium.  At a shear of $\pi/2$,
however, the current structure at the null becomes only a few grid
cells wide and reconnection begins.  For example, in Fig. 4d some of
the red and blue lines clearly have reconnected and joined the green
flux system.  The simulation could find no equilibrium for this
shear. When we tried to let the system relax by continuing the
simulation beyond the shearing phase, the reconnection became
progressively stronger with magnetic islands appearing at the
interface between the red and blue systems. The velocities there
became very large and the density plummetted, halting the simulation.

Although our code could not simulate the actual eruption, it did prove
the key result that the energy of the sheared MHD field greatly
exceeds $E_{min}$. In Figure 3 we compare the magnetic energies as a
function of shear for the force-free and MHD solutions. One may
question the validity of such a comparison because the MHD simulation
also includes the effects of the plasma energy, however, the change in
plasma thermal energy during the whole shearing phase (before
appreciable reconnection) was less than 10\% of the magnetic energy
change.  It is evident from Fig. 3 that the MHD solution energy
exceeds $E_{min}$ even for a shear of $\pi/4$, while the free energy
of the MHD solution is approximately twice that of $E_{min}$ for a
shear of $3\pi/8$.  Therefore, there is no way for the system to
evolve to its minimum energy state except by transferring this excess
energy to the plasma, i.e. by a violent eruption.

\section{Discussion}

The most important result of the MHD simulation is that contrary to
many other calculations, current sheets do not form at the separatrix
until late in the shearing. There are two reasons for this
result. First, no shear is applied at the separatrix. This is the main
difference between this simulation and other simulations, including
our own ({\it e.g.,} \cite{Karpen96,Karpen98}). If a shear were to be applied
at the separatrix of our configuration, then a discontinuous
$B_{\phi}$ would be created due to the discontinuity in relative
footpoint positions, and a current sheet would immediately appear,
just as in the previous simulations.  But we apply a shear only near
the neutral line, far from the separatrix. In fact, the field outside
the shear region should always be nearly current-free for our 2.5D
system since a quasi-static force-free approximation should be valid
everywhere except near the null, and an unsheared 2.5D force-free
field must be potential. This is exactly what our simulation finds,
the current is completely negligible except in the shear region and
in the high-$\beta$ region around the null.

Second, discontinuities do not appear at the null, itself, because we
include in our model a finite plasma pressure so that the system is
far from force-free near the null. In order to calculate the evolution
of the null in the first place, the plasma pressure and dynamics must
be fully included.  Without the plasma the null could not even change
its position. The null moves outward only because stress is
transmitted from the field to the plasma in the low-$\beta$ region,
imparting momentum to the plasma, which then moves the field in the
high-$\beta$ region around the null.  The plasma pressure is also
essential for a smooth equilibrium to form. As the inner flux system
expands outward it compresses the plasma in the high-$\beta$ region
around the null, increasing the plasma pressure there until force
balance is achieved, as in the simple analytic model described above.
It is only when the null region is squeezed down to the dissipation
length scale, which corresponds to a few grid points in our numerical
system, that significant reconnection begins and equilibrium becomes
impossible to maintain. Of course, on the Sun the magnetic Reynolds
number is very large, and the dissipation scale is many orders of
magnitude smaller than global scale of the magnetic field.

An interesting implication of this discussion is that the exact energy
equation for the plasma must play a crucial role in determining when
eruption will occur. In the simulation we assumed a simple adiabatic
energy equation (8), but radiation losses, thermal conduction, and
ohmic heating may all be important.  If the plasma at the null cools
as it compresses, then the width of the null region can decrease
rapidly to the dissipation scale. On the other hand, if the plasma
heats then the null region resists compression and reconnection will
be delayed. Hence, we have the very interesting situation that the
dynamics of huge phenomena such as CMEs may be controlled by detailed
plasma processes occurring in relatively tiny regions.

Another interesting property of our ``magnetic breakout'' model for CME
initiation/flare triggering is that it involves several observable
features which could form the basis for an effective prediction
scheme.  One feature is that the shear is concentrated near the
neutral line, in agreement with observations (\cite{Schmieder96}) and
with our theory for prominence formation (\cite{Antiochos94}). Note
that the shear responsible for eruptions cannot be due to differential
rotation since that would preferentially shear long, high-lying flux
rather than the short, low-lying flux near the neutral
line. Differential rotation would produce the opposite shear profile
from that shown in Figure 2.  The upcoming Solar-B mission should be
able to measure the magnetic shear at the photosphere with great
accuracy, and determine the relation, if any, between shear
distribution and eruptive activity.

Another feature of the model is that multiple flux systems must be
present in order for an eruption to occur.  It is well known that
large eruptive flares only occur in complex topologies, in particular
$\delta$-spot regions. It can be shown that a $\delta$-spot has a 3D
magnetic topology of a four-flux system, exactly analogous to the 2.5D
model discussed here (\cite{Antiochos98}).  We propose that this
explains why $\delta$-spots are so flare productive, whereas bipolar
spot regions are not. Note that this provides a clear distinction
between our breakout model and the usual tether-cutting model
(e.g. \cite{Sturrock89,Moore92}). Since the tether-cutting model
involves only a single arcade, it would predict that a strongly
sheared bipolar spot region is just as likely to flare as a
$\delta$-spot region.

CMEs also favor magnetic complexity.  Both Mauna Loa and LASCO
observations indicate that CMEs are more common in regions where the
coronal field has a multipolar structure, rather than a simple bipolar
structure (\cite{McAllister95,Schwenn96}).  In addition, observations
suggest that many CMEs involve more than one neutral line and,
therefore, more than one flux region (\cite{Webb97}).

On the other hand, there are numerous examples of high-latitude CMEs
that appear to involve only a bipolar magnetic topology.  It would
seem that this type of event contradicts the breakout model.  However,
the high-latitude CMEs not associated with active regions are
invariably slow, with height-time profiles similar to that of the slow
solar wind rather than flare-associated ejection ({\it e.g.}
\cite{Sheeley97}).  We propose that this type of CME should not be
considered a coronal mass {\bf ejection}, but merely a coronal mass
{\bf expansion}.  The key point is that, if the magnetic field expands
outward slowly due to shear while the plasma maintains its temperature
due to coronal heating, then at a sufficiently large height (a few
solar radii), the plasma will begin to dominate the field and expand
outward indefinitely as predicted by the Parker solar wind model. Most
CMEs may be just this type of pressure driven expansion.  But obviously
this process cannot explain the fast CMEs, which appear in the
coronagraph field of view with velocities much faster than the wind,
and may even slow down as they travel upward from the low corona
(\cite{Sheeley97}). We propose that fast CMEs must be due to magnetic
breakout, most likely occurring in strong active region fields.

From the viewpoint of prediction the most important feature of our
model is the reconnection that must occur above the erupting
arcade. This reconnection is not expected to release much energy,
since most of the free magnetic energy is stored in the low-lying
sheared field. Note also that the reconnection occurs on long field
lines far from any neutral line. Therefore, it is unlikely to produce
significant heating or strong X-ray emission, but it may be detectable
in radio/microwave emission from nonthermal particles accelerated
by this reconnection. We believe that radio and microwave observations
will provide the best test of the magnetic breakout model for CME
initiation.

\acknowledgments

This work has been supported in part by NASA and ONR.

\vfill\eject
%\clearpage

\begin{figure}
\caption{ {\bf (a)} Initial potential magnetic field. The
field is symmetric about the axis of rotation and the equator, so only
one quadrant is shown. The photospheric boundary surface is indicated
by the light grey grid. Magnetic field lines are colored (red, green,
or blue) according to their flux system. Two types of blue field lines
are indicated, higher-lying light-blue unsheared field and low-lying
dark-blue field which is sheared later in the simulation.
{\bf (b)} The force-free field after a shear of $\pi/8$. The 
field lines shown correspond to those in 1a. They are traced from the same
footpoint position on the photosphere as in 1a.
{\bf (c)} As above, but for a shear of $3\pi/8$.
{\bf (d)} As above, but for a shear of $\pi/2$. \label{fig1}}
\end{figure}

\begin{figure}
\caption{ The normalized shear profile as a function of colatitude. \label{fig2}}
\end{figure}

\begin{figure}
\caption{ The magnetic free energy as a function of shear for the 
force-free field solution (lower curve) and the MHD solution (upper curve).
The shear is in units of $\pi$ and the free energy is expressed as percentage 
of the initial, potential field energy. \label{fig3}}
\end{figure}

\begin{figure}
\caption{ The  MHD solution after a shear of $\pi/8$ {\bf (4a)},
$\pi/4$ {\bf (4b)}, $3\pi/8$ {\bf (4c)}, and $\pi/2$ {\bf (4d)}.
The field lines shown are the same as those in Figure 1. \label{fig4}}
\end{figure}

\end{document}